\documentclass[aps]{revtex4}
\usepackage{amssymb}
\usepackage[dvips]{graphicx}
\usepackage{amsmath}
\usepackage{graphicx}
\usepackage{makeidx}
\usepackage{subfigure}
\begin{document}

\title{Vortical light bullets in second-harmonic-generating media supported
by a trapping potential}
\author{Hidetsugu Sakaguchi$^{1}$ and Boris A. Malomed$^{2}$}
\address{$^{1}$Department of Applied Science for Electronics and Materials,\\
Interdisciplinary Graduate School of Engineering Sciences,\\
Kyushu University, Kasuga, Fukuoka 816-8580, Japan\\
\email{^{}sakaguchi@asem.kyushu-u.ac.jp}\\
$^{2}$Department of Physical Electronics, School of Electrical Engineering,\\
Faculty of Engineering, Tel Aviv University, Tel Aviv 69978, Israel}

\begin{abstract}
\begin{center}{\bf Abstract}
\end{center}

We introduce a three-dimensional (3D) model of optical media with
the quadratic ($\chi ^{(2)}$) nonlinearity and an effective 2D
isotropic harmonic-oscillator (HO) potential. While it is well known
that 3D $\chi ^{(2)}$ solitons with embedded vorticity (``vortical
light bullets") are unstable in the free space, we demonstrate that
they have a broad stability region in the present model, being
supported by the HO potential against the splitting instability. The
shape of the vortical solitons may be accurately predicted by the
variational approximation (VA). They exist above a threshold value
of the total energy (norm) and below another critical value, which
determines a stability boundary. The existence threshold vanishes is
a part of the parameter space, depending on the mismatch parameter,
which is explained by means of the comparison with the 2D
counterpart of the system. Above the stability boundary, the vortex
features shape oscillations, periodically breaking its axisymmetric
form and restoring it. Collisions between vortices moving in the
longitudinal direction are studied too. The collision is strongly
inelastic at relatively small values of the velocities, breaking the
two colliding vortices into three, with the same vorticity. The
results suggest a possibility of the creation of \emph{stable} 3D
optical solitons with the intrinsic vorticity.

\end{abstract}

\maketitle

\section{Introduction}

\noindent The second-harmonic-generating ($\chi ^{(2)}$, i.e.,
quadratic) nonlinearity is one of basic features of photonic media.
Among other fundamental effects, the $\chi ^{(2)}$ interactions give
rise to two-color solitons of diverse types
\cite{review0,review1,review2,review}, \cite{KA}. In particular, the
use of this nonlinearity is critically important for the creation of
stable two- and three-dimensional (2D and 3D) solitons, because, in
contrast to the cubic ($\chi ^{(3)}$) self-focusing, the $\chi
^{(2)}$ interaction between the fundamental-frequency (FF) and
second-harmonic (SH) fields does not induce the wave collapse
\cite{Rubenchik}, which destabilizes multidimensional solitons in
$\chi ^{(3)}$ settings \cite{review,Romania}. Due to this
circumstance, the $\chi ^{(2)}$ solitons were originally
created, in the spatial domain, as stable (2+1)-dimensional beams \cite%
{first-exper}. The possibility of the creation of fully localized 3D
spatiotemporal solitons is suggested too by the absence of the collapse \cite%
{Drummond,add1,add2,add3}. This has not been achieved yet in experiments,
the best result being a spatiotemporal soliton which is self-trapped in the
longitudinal and one transverse directions, while the confinement along the
other transverse coordinate was provided by the waveguiding structure \cite%
{Wise1,Wise2}. The $\chi ^{(2)}$ nonlinearity is also promising for the
creation of ultrashort few-cycle optical solitons \cite{ultra},

The creation of self-trapped beams with intrinsic vorticity is
another natural possibility in the 2D setting, with the SH and FF
fields carrying topological charges $m$ and $m/2$, respectively. For
even $m$, these modes are classified as solitary vortices with
topological charge $m/2$. States with odd values of $m$ are not
possible, as the topological charge of the FF component cannot take
half-integer values (vortices with a half-integer optical angular
momentum, in the form of mixed screw-edge dislocations, can be
created by passing the holding beam through a spiral-phase plate
displaced off the beam's axis \cite{half}). Unlike the fundamental
$\chi ^{(2)}$ solitons with $m=0$, solitary vortices in the free
space are always unstable against azimuthal perturbations, which
split them into sets of separating segments, as shown both
theoretically \cite{splitting1,splitting2,splitting4,splitting5} and
experimentally \cite{splitting-exp}. A similar splitting instability
of vortices was predicted in the framework of the three-wave (alias
Type-II) $\chi ^{(2)}$ system, which includes two components of the
FF field \cite{splitting-3W,splitting-3W2}.

Vortical solitons are well known too as solutions to the 2D and 3D
nonlinear Schr\"{o}dinger (NLS) equation with the cubic ($\chi
^{(3)}$) self-focusing nonlinearity \cite{Kruglov}. They are also
unstable against azimuthal perturbations, which develops faster than
the above-mentioned collapse, i.e., the vortex splits into
fragments, which later suffer the intrinsic collapse \cite{review}.
A method for the stabilization of the vortices was elaborated in the
framework of the 2D and 3D Gross-Pitaevskii (GP) equations for
atomic Bose-Einstein condensates (BECs) with attractive inter-atomic
interactions, which are similar to the NLS equations for photonic
media with the self-focusing $\chi ^{(3)}$ nonlinearity \cite{BEC}.
It has been shown in detail theoretically
\cite{cubic-in-trap1,cubic-in-trap2,cubic-in-trap3,cubic-in-trap4,cubic-in-trap5,cubic-in-trap6,cubic-in-trap7}
that the axisymmetric harmonic-oscillator (HO) potential stabilizes
the fundamental solitons in their whole existence domain, and
vortices with topological charge $1$ in a part of the region where
they exist.

In optical media, the effective 2D trapping potential can be
realized too, in the form of an appropriate transverse profile of
the local refractive index \cite{KA}. The stabilization of 2D vortex
solitons by means of the isotropic HO potential in the medium with
the $\chi ^{(2)}$ nonlinearity was reported in our recent work
\cite{we}. A 2D photonic crystal can be used to build this setting,
with the $\chi ^{(2)}$ nonlinearity provided by the poled liquid
\cite{Du} or solid \cite{Luan} material filling its voids. In fact,
the profile of the effective potential does not need to be exactly
parabolic (HO), due to the strong localization of the trapped modes.
Our recent analysis \cite{we} demonstrates the stabilization of both
two-color $\chi ^{(2)}$ vortices and single-color ones, only the SH
component being present in the latter case. Stability regions were
found for both types of the modes. In particular, while the
single-color states are actually linear modes supported by the
trapping potential, their stability is determined by the $\chi
^{(2)}$ interactions with small perturbations in the FF field.
Two-color trapped vortices exist above a finite threshold values,
which is identical to the instability boundary of the single-color
vortex mode.

A natural extension of the previous analysis, which is the subject
of the present work, is to consider a possibility of the
stabilization of two-color 3D vortex solitons (``vortical light
bullets" \cite{review2}) by the 2D axisymmetric trapping potential,
which may open the way to creating such self-trapped spatiotemporal
modes. Thus far, they evaded attempts of the experimental
observation (it is obvious that, unlike the 2D setting, single-color
localized modes cannot exist in the 3D $\chi ^{(2)}$ system with the
2D potential). The 3D model, based on the system of coupled
equations for the FF and SH fields, is introduced in Section II. In
particular, an essential parameter, specific to the 3D model, is the
ratio of the group-velocity-dispersion (GVD) coefficients for the FF
and SH waves. In fact, essentially the same system of GP equations
for atomic and molecular wave functions, applies to the BEC in
atomic-molecular mixtures, where the role of the GVD ratio is played
by the ratio of the atomic and molecular masses
\cite{BEC0,BEC1,BEC2,BEC4}, hence the mechanism of the stabilization
of the 3D vortex solitons trapped in the 2D HO potential may be
implemented in the BEC mixture too. In Section II we also formulate
the variational approximation (VA) for stationary states.

Stationary solutions for the trapped 3D vortical modes are considered in
Section III. The results are obtained by means of numerical methods, and in
a semi-analytical form based on the VA. In those cases when the underlying
ansatz is appropriate, the accuracy of the VA is very good, in comparison
with numerical results. Depending on the mismatch parameter, the vortex
modes may exist above a finite threshold value of the norm (total energy),
which can be explained in a simple form. The stability of the vortex
solitons is considered, by means of systematic simulations of the perturbed
evolution, in Section IV. The vortex modes are stable below a certain
critical value of the norm. Above the instability boundary, the vortex
develop oscillations, maintaining its vorticity and demonstrating periodic
breaking and recovery of the axial symmetry of its shape. Collisions between
vortex solitons moving in the longitudinal direction are also considered in
Section IV, a noteworthy result being that, at sufficiently small
velocities, two colliding vortices may break up into three. The paper is
concluded by Section V.

\section{The model and the variational approximation}

\noindent The equations for the evolution of the FF and SH field
amplitudes, $u\left( x,y,z,\tau \right) $ and $v\left( x,y,z,\tau
\right) $, in the 3D setting, with propagation distance $z$,
transverse coordinates $\left( x,y\right) $,
and reduced time $\tau \equiv t-z/V$ are taken in the known form \cite%
{review0,review1,review2} $\ $($V$ is the common group velocity of
the FF and SH waves, assuming that the group-velocity mismatch
between them may be made equal to zero by means of an appropriate
Galilean boost):
\begin{eqnarray}
i\frac{\partial u}{\partial z}+\frac{1}{2}\left( \frac{\partial ^{2}}{%
\partial x^{2}}+\frac{\partial ^{2}}{\partial y^{2}}+\frac{\partial ^{2}}{%
\partial \tau ^{2}}\right) u+u^{\ast }v-U(x,y)u &=&0,  \notag \\
2i\frac{\partial v}{\partial z}+\frac{1}{2}\left( \frac{\partial ^{2}}{%
\partial x^{2}}+\frac{\partial ^{2}}{\partial y^{2}}+D\frac{\partial ^{2}}{%
\partial \tau ^{2}}\right) v-qv+\frac{1}{2}u^{2}-4U(x,y)v &=&0,  \label{eqs}
\end{eqnarray}%
where the asterisk stands for the complex conjugate, $q$ is the real
mismatch coefficient, the anomalous-GVD coefficient at the FF is normalized
to be $1$, and $D$ is the ratio of the SH and FF GVD coefficients. In the
case of $D=1$, the diffraction-dispersion operators for the FF and SH
components feature the spatiotemporal isotropy \cite{Drummond}. As said
above, the axisymmetric modulation of the refractive index is represented by
a spatially-isotropic HO potential, $U(x,y)=\left( \Omega ^{2}/2\right)
\left( x^{2}+y^{2}\right) $. By means of a further rescaling, we fix $\Omega
=1/2$, while $q$ is kept as a free constant. Stationary modes are
characterized by their total energy (norm),
\begin{equation}
N=\int \int \int \left( |u|^{2}+4|v|^{2}\right) dxdyd\tau ,  \label{N0}
\end{equation}%
which is dynamical invariant of the system. Three other invariants are the
angular momentum:%
\begin{equation}
M=\int \int \int \text{\textrm{Im}}\left\{ u^{\ast }\left(
yu_{x}-xu_{y}\right) \right\} dxdyd\tau ,  \label{M}
\end{equation}%
the linear momentum in the $\tau $ direction:%
\begin{equation}
P=\int \int \int \text{\textrm{Im}}\left\{ u^{\ast }u_{\tau }\right\}
dxdyd\tau ,  \label{P}
\end{equation}%
and the Hamiltonian:%
\begin{eqnarray}
H &=&\int \int \int \left\{ \left[ \frac{1}{2}\left(
|u_{x}|^{2}+|u_{y}|^{2}+|u_{\tau }|^{2}+|v_{x}|^{2}+|v_{y}|^{2}+D|v_{\tau
}|^{2}\right) \right] \right.  \notag \\
&&\left. +U(r)\left( |u|^{2}+4|v|^{2}\right) +q|v|^{2}-\frac{1}{2}\left(
u^{2}v^{\ast }+u^{2\ast }v\right) \right\} dxdyd\tau .  \label{H}
\end{eqnarray}

Equations (\ref{eqs}) can be derived from the corresponding action, $S=\int
Ldz$, with Lagrangian
\begin{eqnarray}
L &=&\int \int \int \left\{ \left[ iu_{z}u^{\ast }+2iv_{z}v^{\ast }-\frac{1}{%
2}\left( |u_{x}|^{2}+|u_{y}|^{2}+|u_{\tau
}|^{2}+|v_{x}|^{2}+|v_{y}|^{2}+D|v_{\tau }|^{2}\right) \right] \right.
\notag \\
&&\left. -U(r)\left( |u|^{2}+4|v|^{2}\right) -q|v|^{2}+\frac{1}{2}\left(
u^{2}v^{\ast }+u^{2\ast }v\right) \right\} dxdyd\tau .  \label{L}
\end{eqnarray}%
This fact may be used to develop the VA for spatiotemporal vortical solitons
with propagation constant $k$, based on the following natural ansatz, which
corresponds to topological charge $1$ and is written in terms of polar
coordinates $\left( r,\theta \right) $ in the plane of $\left( x,y\right) $:
\begin{equation}
u=\frac{u_{0}r\exp \left( ikz-\alpha _{1}r^{2}+i\theta \right) }{\cosh
\left( \beta \tau \right) },\;v=\frac{v_{0}r^{2}\exp \left( ikz-\alpha
_{2}r^{2}+2i\theta \right) }{\cosh \left( \beta \tau \right) }.
\label{ansatz}
\end{equation}%
Here $\left( u_{0},v_{0}\right) $ and $\alpha _{1,2}^{-1/2}$ are,
respectively, the amplitudes and radial widths of the two components, and $%
\beta ^{-1}$ is their common temporal width. The norm (\ref{N0}) of the
ansatz is%
\begin{equation}
N=2\pi \int_{-\infty }^{+\infty }d\tau \int_{0}^{\infty
}rdr(|u|^{2}+4|v|^{2})=\frac{4\pi }{\beta }\left( \frac{u_{0}^{2}}{8\alpha
_{1}^{3}}+\frac{v_{0}^{2}}{2\alpha _{2}^{3}}\right)  \label{N}
\end{equation}

The substitution of ansatz (\ref{ansatz})\ into the Lagrangian yields
\begin{gather}
\frac{L}{2\pi }=-\frac{kN}{2\pi }-\frac{u_{0}^{2}}{2\alpha _{1}\beta }-\frac{%
3v_{0}^{2}}{4\alpha _{2}\beta }-\frac{u_{0}^{2}\beta }{24\alpha _{1}^{2}}-%
\frac{Dv_{0}^{2}\beta }{24\alpha _{2}^{3}}  \notag \\
-\frac{\Omega ^{2}u_{0}^{2}}{8\alpha _{1}^{3}\beta }-\frac{3\Omega
^{2}v_{0}^{2}}{4\alpha _{2}^{4}\beta }-\frac{qv_{0}^{2}}{4\alpha
_{2}^{3}\beta }+\frac{\pi u_{0}^{2}v_{0}}{2\beta (2\alpha _{1}+\alpha
_{2})^{3}}.  \label{LL}
\end{gather}%
Using the relation following from Eq. (\ref{N}), $v_{0}^{2}=\left[
N_{0}\beta -u_{0}^{2}/(4\alpha _{1}^{2})\right] \alpha _{2}^{3}$, with $%
N_{0}\equiv N/(2\pi )$, Lagrangian~(\ref{LL}) is cast into the form in which
$N_{0}$ may be considered as a given constant, while $u_{0},\alpha
_{1},\alpha _{2}$, and $\beta $ are treated as variational parameters:
\begin{eqnarray}
\frac{L}{2\pi } &=&-kN_{0}-\frac{u_{0}^{2}}{2\alpha _{1}\beta }-\frac{%
3\alpha _{2}}{4\beta }\left( N_{0}\beta -\frac{u_{0}^{2}}{4\alpha _{1}^{2}}%
\right) -\frac{u_{0}^{2}\beta }{24\alpha _{1}^{2}}-\frac{D\beta }{24}\left(
N_{0}\beta -\frac{u_{0}^{2}}{4\alpha _{1}^{2}}\right) -\frac{\Omega
^{2}u_{0}^{2}}{8\alpha _{1}^{3}\beta }  \notag \\
&&-\left( \frac{3\Omega ^{2}}{4\alpha _{2}\beta }+\frac{q}{4\beta }\right)
\left( N_{0}\beta -\frac{u_{0}^{2}}{4\alpha _{1}^{2}}\right) +\frac{\pi
u_{0}^{2}\alpha _{2}^{3/2}}{2\beta (2\alpha _{1}+\alpha _{2})^{3}}\left(
N_{0}\beta -\frac{u_{0}^{2}}{4\alpha _{1}^{2}}\right) ^{1/2}.  \label{LLL}
\end{eqnarray}%
The system of variational equations produced by the latter Lagrangian, $%
\partial L/\partial u_{0}=\partial L/\partial \alpha _{1}=\partial
L/\partial \alpha _{2}=\partial L/\partial \beta =0$, take a cumbersome
form, which, nevertheless, admits a numerical solution:
\begin{gather}
-\frac{u_{0}}{\alpha _{1}\beta }+\frac{3u_{0}\alpha _{2}}{8\beta \alpha
_{1}^{2}}-\frac{u_{0}\beta }{12\alpha _{1}^{2}}+\frac{D\beta u_{0}}{48\alpha
_{1}^{2}}-\frac{\Omega ^{2}u_{0}}{4\alpha _{1}^{3}\beta }+\frac{3\Omega
^{2}u_{0}}{8\alpha _{2}\beta \alpha _{1}^{2}}+\frac{qu_{0}}{8\beta \alpha
_{1}^{2}}  \notag \\
+\frac{\pi u_{0}\alpha _{2}^{3/2}}{\beta (2\alpha _{1}+\alpha _{2})^{3}}%
\left( N_{0}\beta -\frac{u_{0}^{2}}{4\alpha _{1}^{2}}\right) ^{1/2}-\frac{%
\pi u_{0}^{3}\alpha _{2}^{3/2}}{8\beta \alpha _{1}^{2}(2\alpha _{1}+\alpha
_{2})^{3}}\left( N_{0}\beta -u_{0}^{2}/(4\alpha _{1}^{2})\right) ^{-1/2}=0,
\notag \\
\frac{u_{0}^{2}}{2\alpha _{1}^{2}\beta }-\left( \frac{3\alpha _{2}}{2\beta }+%
\frac{D\beta }{12}+\frac{3\Omega ^{2}}{2\alpha _{2}\beta }\right) \frac{%
u_{0}^{2}}{4\alpha _{1}^{3}}+\frac{u_{0}^{2}\beta }{12\alpha _{1}^{3}}+\frac{%
3\Omega ^{2}u_{0}^{2}}{8\alpha _{1}^{4}\beta }  \notag \\
-\frac{3\pi u_{0}^{2}\alpha _{2}^{3/2}}{\beta (2\alpha _{1}+\alpha _{2})^{4}}%
\left( N_{0}\beta -\frac{u_{0}^{2}}{4\alpha _{1}^{2}}\right) ^{1/2}+\frac{%
\pi u_{0}^{4}}{8\beta \alpha _{1}^{3}(2\alpha _{1}+\alpha _{2})^{3}}\left(
N_{0}\beta -\frac{u_{0}^{2}}{4\alpha _{1}}\right) ^{-1/2}=0,  \notag \\
\left( -\frac{3}{4\beta }+\frac{3\Omega ^{2}}{4\alpha _{2}^{2}\beta }\right)
\left( N_{0}\beta -\frac{u_{0}^{2}}{4\alpha _{1}^{2}}\right)  \notag \\
+\left( -\frac{3\pi u_{0}^{2}\alpha _{2}^{3/2}}{2\beta (2\alpha _{1}+\alpha
_{2})^{4}}+\frac{3\pi u_{0}^{2}\alpha _{2}^{1/2}}{4\beta (2\alpha
_{1}+\alpha _{2})^{3}}\right) \left( N_{0}\beta -\frac{u_{0}^{2}}{4\alpha
_{1}^{2}}\right) ^{1/2}=0,  \notag \\
\frac{u_{0}^{2}}{2\alpha _{1}\beta ^{2}}-\left( \frac{3\alpha _{2}}{2\beta }+%
\frac{D\beta }{12}+\frac{3\Omega ^{2}}{2\alpha _{2}\beta }\right) \frac{N_{0}%
}{2}+\left( \frac{3\alpha _{2}}{4\beta }-\frac{D}{24}+\frac{3\Omega ^{2}}{%
4\alpha _{2}\beta ^{2}}\right) \left( N_{0}\beta -\frac{u_{0}^{2}}{4\alpha
_{1}^{2}}\right)  \notag \\
-\frac{u_{0}^{2}}{24\alpha _{1}^{2}}+\frac{\Omega ^{2}u_{0}^{2}}{8\alpha
_{1}^{3}\beta ^{2}}-\frac{qu_{0}^{2}}{16\beta ^{2}\alpha _{1}^{2}}-  \notag
\\
\frac{\pi u_{0}^{2}\alpha _{1}^{3/2}}{2\beta ^{2}(2\alpha _{1}+\alpha
_{2})^{3}}\left( N_{0}\beta -\frac{u_{0}^{2}}{4\alpha _{1}^{2}}\right)
^{1/2}+\frac{\pi u_{0}^{2}N_{0}\alpha _{2}^{3/2}}{4\beta (2\alpha
_{1}+\alpha _{2})^{3}}\left( N_{0}\beta -\frac{u_{0}^{2}}{4\alpha _{1}^{2}}%
\right) ^{-1/2}=0.  \label{VA}
\end{gather}

\section{Numerical and semi-analytical results for stationary vortices}

\noindent At first, we numerically solved Eqs.~(\ref{eqs}) by means
of the imaginary-time-evolution method (it was earlier applied to
searching for vortical trapped states in the framework of the GP
equation with the cubic nonlinearity \cite{Tosi}), starting with
initial conditions corresponding to the 3D vortical wave form with
topological charge $1$ such as
\begin{equation}
u(x,y,\tau )=0.5re^{-0.25r^2}e^{-0.5\tau^2}e^{i\theta },\;v(x,y,\tau )=0.
\end{equation}
Shapes of the so generated localized vortical states were compared to the
prediction of the VA, obtained for the same parameters, and the stability of
the states was checked by simulations of their perturbed evolution in the
framework of Eq.~(\ref{eqs}) in real time. Equations (\ref{eqs}) were solved
numerically in the 3D domain defined as $0\leq \left\{ x,y,\tau \right\}
\leq \left\{ L,L,T\right\} ,$ with $L\times L\times T=25\times 25\times 50$,
and periodic boundary conditions. Accordingly, the polar coordinates $\left(
r,\theta \right) $ are defined with respect to the central point, $x=y=L/2$.
This size of the integration domain was sufficient to completely accommodate
all the vortical solitons produced by the numerical solution.

\begin{figure}[tbp]
\begin{center}
\includegraphics[height=3.cm]{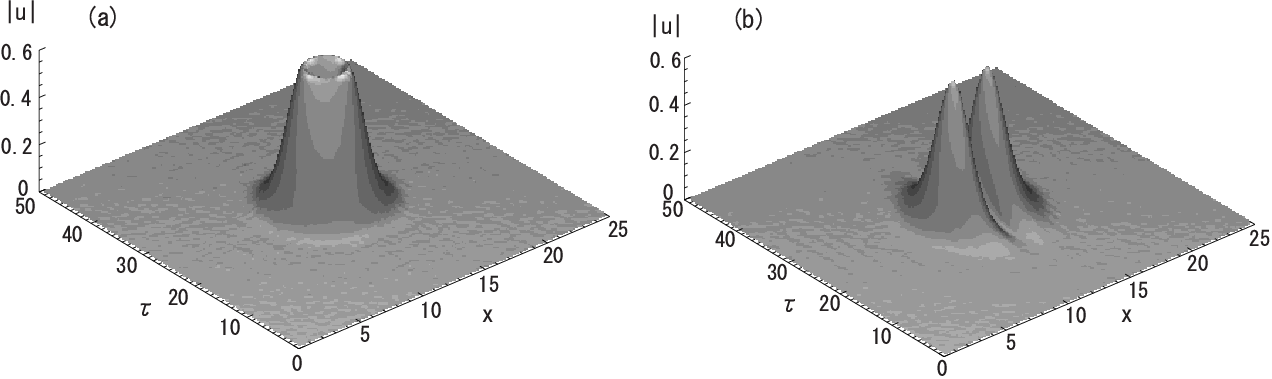}
\end{center}
\caption{Plots of $|u(x,y,\protect\tau )|$ in the cross section drawn
through $\protect\tau =T/2\equiv 25$ (a) and $y=12.5$ (b). }
\label{fig1}
\end{figure}

\begin{figure}[t]
\begin{center}
\includegraphics[height=3.5cm]{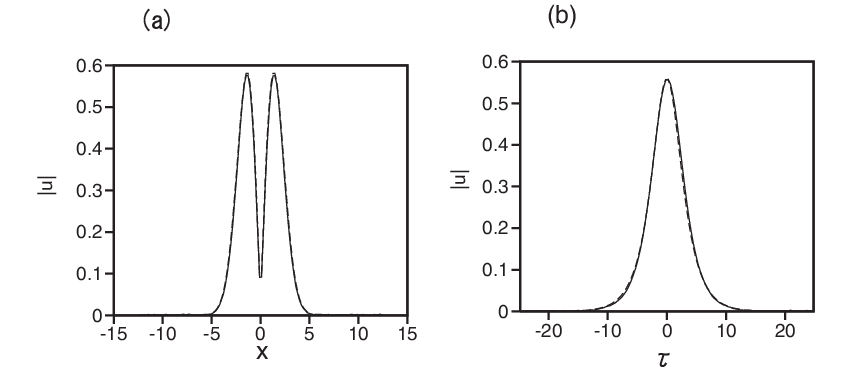}
\end{center}
\caption{(a) The plot of $|u(x-L/2,y,\protect\tau )|$ at $y=L/2,\protect\tau %
=T/2$ obtained from the direct numerical solution (the solid curve) and its
VA-predicted counterpart, $u_{0}|x|\exp \left( -\protect\alpha %
_{1}x^{2}\right) $, with $u_{0}=0.704,\protect\alpha _{1}=0.268$ (the dashed
curve). (b) The same for $|u(x,y,\protect\tau -T/2)|$ at $x=1.663,y=L/2$,
the respective VA-predicted profile being $u_{0}|x|\exp \left( -\protect%
\alpha _{1}x^{2}\right) \mathrm{sech}(\protect\beta \protect\tau )$ with $%
u_{0}=0.704,\protect\alpha _{1}=0.268,\protect\beta =0.438$. }
\label{fig2}
\end{figure}

A characteristic example is presented in Fig. \ref{fig1}, which shows a plot
of $|u(x,y,\tau )|$ in cross sections drawn through $\tau =L_{\tau }/2$ (a)
and $y=L/2$ (b) at $N=40,q=1,D=1,\Omega =0.5$. In particular, Fig. \ref{fig1}%
(a) shows the vortex structure per se, with amplitude $|u|$ vanishing at the
center, and Fig. \ref{fig1}(b) shows the self-trapped structure in the $\tau
$ direction. For the same parameters, the numerical solution of the
variational equations~(\ref{VA}) yields $u_{0}=0.704,~\alpha
_{1}=0.268,~\alpha _{2}=0.507$, and $\beta =0.438$. To compare the full
numerical results with the prediction of the VA, the solid curve in Fig.~\ref%
{fig2}(a) shows the numerically found profile of $|u(x-L/2,y,\tau )|$ at $%
y=L/2,\tau =L_{\tau }/2$, and the dashed curve shows the VA-predicted
counterpart of this profile, taken as per Eq. (\ref{ansatz}), i.e., $%
u_{0}|x|\exp \left( -\alpha _{1}x^{2}\right) $. Further, the solid curve in
Fig.~\ref{fig2}(b) shows the temporal profile, $|u(x,y,\tau -L_{\tau }/2)|$,
at fixed $x=1.663$ and $y=L/2$, and the dashed curve shows its VA-predicted
counterpart, $u_{0}|x|\exp \left( -\alpha _{1}x^{2}\right) /\cosh (\beta
\tau )$ at the same point, $x=1.663$. It is seen that the VA predicts the
numerical results in a practically exact form.
\begin{figure}[t]
\begin{center}
\includegraphics[height=3.5cm]{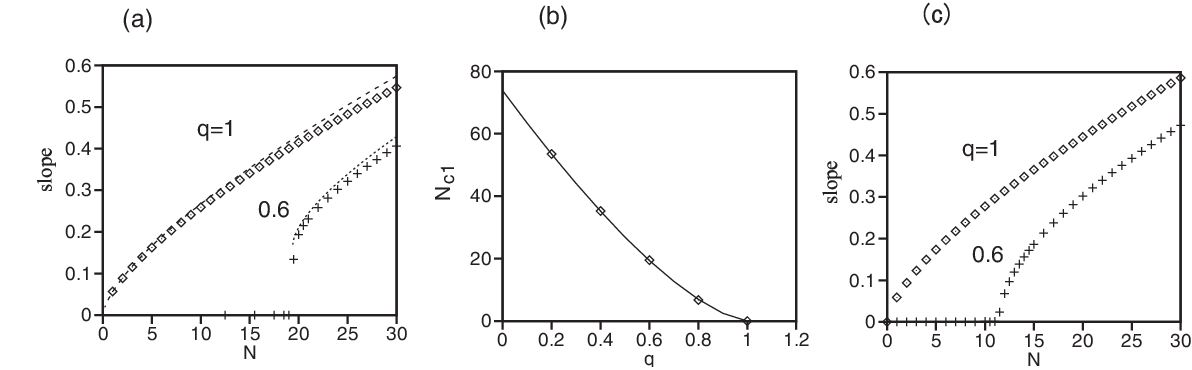}
\end{center}
\caption{(a) The slope of the vortical profile at the center, defined as per
Eq. (\protect\ref{slope}), produced by the numerical results (rhombuses),
and by the VA (dashed curves) at $q=1$ and $0.6$ for $D=1$. (b) The critical
value, $N_{c1}$, as predicted by the VA (the solid curve) and found from the
numerical results (rhombuses), for $D=1$. The vortex solitons do not exist
at $N<N_{c}$. (c) The same as in (a), but for $D=0$ (the case of zero GVD of
the SH\ component). }
\label{fig3}
\end{figure}

Systematic results obtained for the vortex-soliton family in the direct
numerical and VA forms are summarized in Fig. \ref{fig3}, in which panel (a)
shows the slope of the vortical profile at its center,
\begin{equation}
\mathrm{slope~}\equiv r^{-1}|u(r,\tau )|~~\mathrm{at~~}r\rightarrow 0,
\label{slope}
\end{equation}%
as a function of the total energy (norm) $N$, for two values of the
mismatch, $q=1$ and $0.6$. In the latter case, the vortex soliton disappears
discontinuously at a critical value $N=N_{c1}\approx 19.5$, as demonstrated
by the numerical results and the VA alike, and the vortices do not exist at $%
N<N_{c1}$. Further, Fig. \ref{fig3}(b) shows the critical value as a
function of the mismatch. Note that $N_{c1}$ vanishes at $q\geq 1$.

It is relevant to note that a finite threshold value of the norm, $%
N=N_{c1}^{(\mathrm{2D})}$, necessary for the existence of two-color
vortical solitons in the 2D version of Eqs. (\ref{eqs}), was found
in our recent work \cite{we}. As well as in the present model,
$N_{c1}^{(\mathrm{2D})}$ vanishes at $q\geq 1$. The difference from
the 3D case is that stationary 2D single-color vortices trapped in
the HO potential, with the zero FF component (i.e., these are linear
SH modes), exist at $N<N_{c1}$. Obviously, this is not possible in
the 3D case, as the 2D potential cannot trap 3D linear modes. The
vanishing of $N_{c1}^{(\mathrm{2D})}$ at $q\geq 1$ was explained
analytically in our recent work \cite{we}. In fact, the same
explanation applies here, as the problem of the presence or absence
of the finite threshold reduces to the
existence or nonexistence of the solitons with a vanishingly small norm, $%
N\rightarrow 0$. Obviously, in this limit the radial width of the
mode trapped by the HO potential remains finite in the $\left(
x,y\right) $ plane, while the temporal width (in the $\tau $
direction) becomes indefinitely large. Thus, the 3D problem for
$N\rightarrow 0$ actually reduces to its 2D counterpart, for which
an explanation of the absence of the threshold at $q\geq 1$ was
recently given in work \cite{we}.

We have also studied vortex solitons in the spatiotemporally anisotropic
system, with $D<1$ in Eqs. (\ref{eqs}) (usually, the anomalous GVD is weaker
at the SH \cite{review}, therefore it makes sense to consider the case of $%
D<1$). In particular, Fig. \ref{fig3}(c) shows, for this case, the vortex'
slope at the center, defined in Eq. (\ref{slope}), as a function of $N$ at $%
q=1$ and $0.6$ for the limit case of $D=0$ (zero GVD at the SH component).
Note that the vortex soliton disappears continuously both for $q=0.6$ and $1$
at $D=0$ [i.e., $N_{c1}(q=0.6,D=0)=0$], but it disappears by a jump at $q=0$%
, i.e., $N_{c1}(q=0,D=0)$ is finite.

The VA approximation poorly agrees with the numerical results for $D=0$,
which is explained by the fact that ansatz (\ref{ansatz}), postulating equal
temporal widths of the FF and SH components of the vortex soliton, is not
appropriate in this case. For $D<0$, vortex-soliton solutions could not be
found, in agreement with the known general principle that they do not exist
in this case \cite{Isaac,review}.

\begin{figure}[t]
\begin{center}
\includegraphics[height=3.5cm]{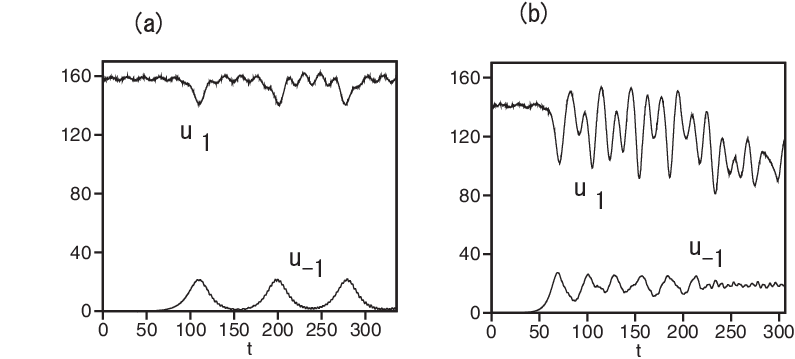}
\end{center}
\caption{(a) The perturbed evolution of the Fourier amplitudes $u_{1}$ and $%
u_{-1}$, defined as per Eq. (\protect\ref{+-}), for $D=1$. (b) The same for $%
D=0$. }
\label{fig4}
\end{figure}
\begin{figure}[tbp]
\begin{center}
\includegraphics[height=3.5cm]{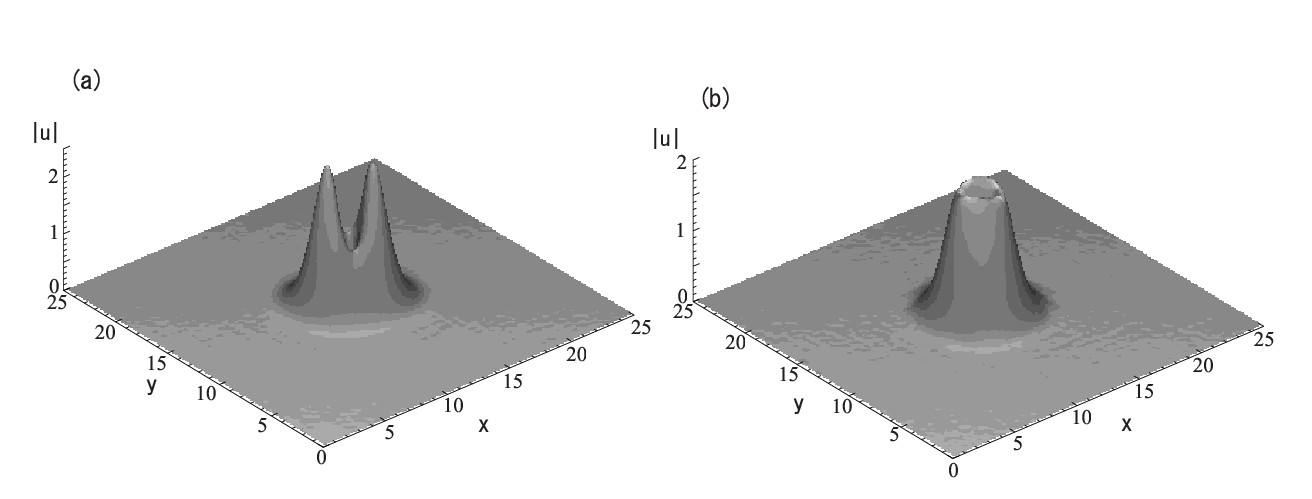}
\end{center}
\caption{Plots of $|u\left( x,y\right) |$ in the cross section drawn through
$\protect\tau =T/2$ at $z=110$ (a) and $z=150$ (b), for the oscillatory
unstable vortex soliton with $N=185$.}
\label{fig5}
\end{figure}

\section{Stability and dynamics of the vortex solitons}

\noindent Systematic simulations demonstrate that the vortex
solitons found above are stable in a large part of their existence
domain. However, they are subject to an oscillatory instability at
$N>N_{c2}$, when the total energy (norm) of the soliton exceeds the
respective critical value, $N_{c2}$. An example of such an
instability is presented in Fig. \ref{fig4}(a), which shows the
evolution of two azimuthal Fourier amplitudes, which are defined as
\begin{equation}
u_{\pm 1}(z)=\left\vert \int \int \int u(x,y,z,\tau )e^{\mp i\theta
}dxdyd\tau \right\vert ,  \label{+-}
\end{equation}%
at $q=1,D=1$ and $N=185$ (obviously, for the stationary vortex $u_{+1}=%
\mathrm{const}$, and $u_{-1}=0$). In the course of the unstable evolution,
amplitude $u_{-1}(z)$ increases from zero, exhibiting oscillatory dynamics.
Further, Fig. \ref{fig4}(b) shows the perturbed evolution of the same
amplitudes for $q=1$, $N=185$, and $D=0$ (zero GVD at the SH component). The
difference from the picture shown in \ref{fig4}(a) is that, at $D=0$, the
oscillations are apparently chaotic.

Figure \ref{fig5} demonstrates that, in terms of the shape of the same
vortex soliton whose evolution is illustrated by Fig. \ref{fig4}(a), the
oscillatory perturbations induce an azimuthal modulational instability of
the axisymmetric shape. The axial symmetry is partly destroyed and then
restored periodically, as one can see from the comparison of Figs. \ref{fig5}
(a) and (b). In the former panel, the modulation depth attains its maximum
exactly at point $z=110$, where $u_{-1}(z)$ has a maximum in Fig. \ref{fig4}
(a), and the axial symmetry is restored at $z=150$, where $u_{-1}(z)$
vanishes in Fig. \ref{fig4}(a). The period of the periodic breakup and
retrieval of the axial symmetry is $Z\approx 80$, the same as observed in
the oscillations of $u_{-1}(z)$ in Fig. \ref{fig4}(a). This dynamical regime
resembles the ones featuring periodic splittings and recombinations of 2D
vortices, which are trapped in the isotropic HO potential in the systems
with $\chi ^{(2)}$ \cite{we} and self-attractive $\chi ^{(3)}$ \cite%
{cubic-in-trap4} nonlinearities.

\begin{figure}[t]
\begin{center}
\includegraphics[height=3.5cm]{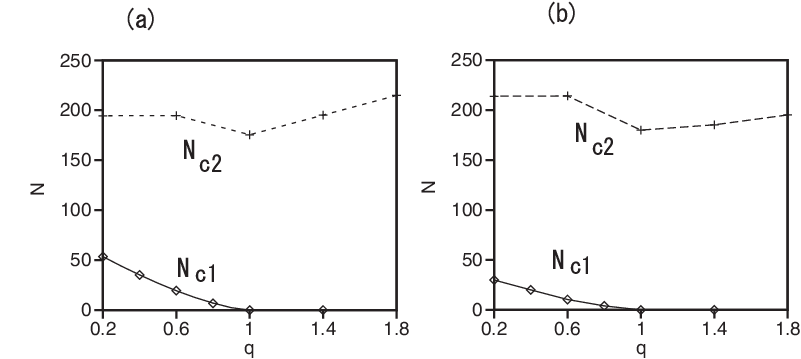}
\end{center}
\caption{The threshold for the onset of the oscillatory instability of the
axisymmetric vortex soliton, shown along with the existence threshold, $%
N_{c1}$, for $D=1$ (a) and $D=0$. Dependence $N_{c1}(q)$ in panel (a) is
tantamount to that shown in Fig. \protect\ref{fig3}(b).}
\label{fig6}
\end{figure}

The critical value for the onset of the oscillatory instability, found from
the direct simulations, is shown in Fig.~\ref{fig6}(a) for several values of
the mismatch, $q=0.2,~0.6,~1.0,~1.4$, and $1.8$, in the systems with the
spatiotemporally isotropic diffraction-dispersion ($D=1$), and with zero GVD
of the SH\ component ($D=0$). Stable vortex solitons exists in the broad
region of $N_{c1}<N<N_{c2}$ [recall $N_{c1}$ is the existence threshold for
the solitons, see Fig. \ref{fig3}(b)]. It is worthy to note that the
stability region is somewhat larger for $D=0$ than for $D=1$.
\begin{figure}[t]
\begin{center}
\includegraphics[height=3.5cm]{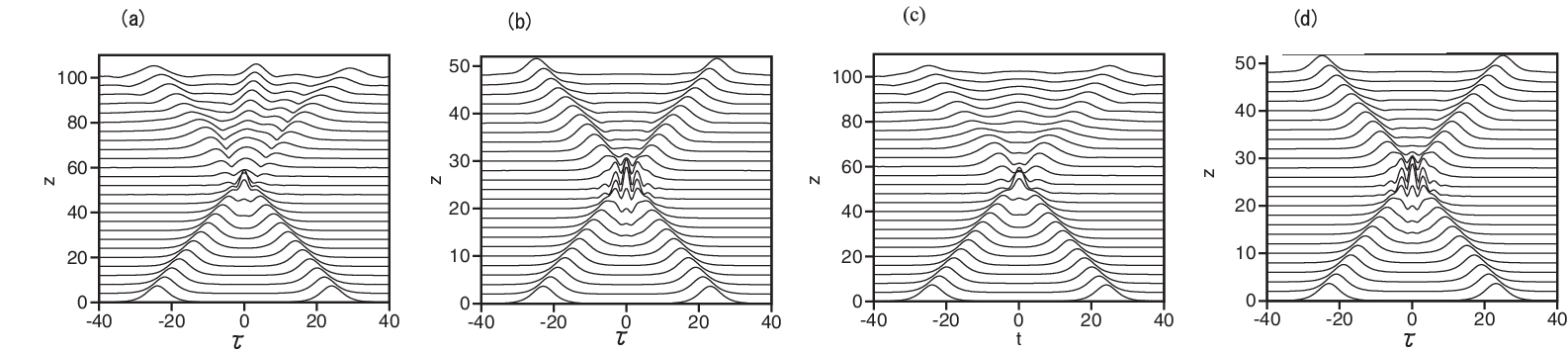}
\end{center}
\caption{Collisions between moving vortex solitons with identical
vorticities, under the action of kicks $\protect\omega =\pm 0.5$ (a);
identical vorticities and $\protect\omega =\pm 1$ (b); opposite vorticities,
$\pm 1$, and kicks $\protect\omega =\pm 0.5$ (c) and $\protect\omega =\pm 1$
(d). Shown are profiles in the cross section drawn in the longitudinal ($%
\protect\tau $) direction through $x-L/2=1.66$ and $y=L/2$. }
\label{fig7}
\end{figure}

The obvious Galilean invariance of Eqs. (\ref{eqs}) with $D=1$ (the case of
the spatiotemporally isotropic diffraction-dispersion operators) in the
longitudinal ($\tau $) direction suggests a possibility to study collisions
between moving solitons, if they are set in motion by means of the
application of properly matched kicks to the FF and SH components, $\left(
u,v\right) \rightarrow \left( ue^{-i\omega \tau },ve^{-2i\omega \tau
}\right) $, where $\omega $ is an arbitrary boost parameter. Figure \ref%
{fig7}(a) shows a typical example of the simulated head-on collision of two
vortex solitons with equal vorticities $1$, kicked by $\omega =\pm 0.5$. The
collision is strongly inelastic, transforming the two colliding vortices
into three, one quiescent and two symmetrically moving ones. We stress that
each soliton produced by this inelastic interaction is a vortex with the
same vorticity, $1$, which was embedded into the initial modes, the total
angular momentum (\ref{M}) and the linear momenetum (\ref{P}) being
conserved. Inelastic collisions between vortices do not generate additional
zero-vorticity solitons. The increase of the kick to $\omega =\pm 1$ results
in a quasi-elastic passage of the colliding solitons through each other, see
Fig. \ref{fig7}(b). Figures \ref{fig7}(c) and (d) demonstrate the collision
of two solitons with opposite vorticities, $+1$ and $-1$ kicked by (c) $%
\omega =\pm 0.5$ and (d) $\omega=\pm 1$.@The collision is inelastic for $%
\omega=\pm 0.5$ and quasi-elastic for $\omega=\pm 1$.

\section{Conclusion}

\noindent The objective of this work is to propose the model of
optical media with the $\chi ^{(2)}$ nonlinearity, in which 3D
vortical solitons, which are always unstable against splitting in
the free space, may be stabilized by means of the effective 2D
isotropic HO (harmonic-oscillator) potential, induced by an
appropriate spatial modulation of the refractive index. A broad
stability region for the 3D vortices has been found, with their
shape accurately predicted by the VA (variational approximation).
The vortices exist with the norm (total energy) exceeding a finite
threshold, which vanishes at values of the mismatch $q\geq 1$. The
latter fact was explained by means of the comparison with similar
results recently reported in the 2D version of the system in our
work \cite{we}. The stability region is limited by another critical
value of the norm, which is much higher than the existence
threshold. Above the stability border, the vortex oscillates,
periodically breaking and restoring its axisymmetric shape.
Collisions between vortices, which may move freely in the
longitudinal (temporal) direction, were also studied. If the
collision velocity is small enough, the collision transforms the two
colliding vortices into three, with the same vorticity. The results
reported in this work suggest that 3D optical vortex solitons may be
created in the stable form.

The analysis can be naturally extended in other directions. In
particular, it may be interesting to study the system with the
Type-II $\chi ^{(2)}$ nonlinearity, i.e., the three-wave system with
two FF components determined by their mutually orthogonal
polarizations \cite{review1,review2}. Further, it is relevant to
consider a possibility to stabilize the 3D vortex solitons using a
periodic (lattice) potential, instead of the HO one, cf. the
analysis performed previously for the $\chi ^{(2)}$
\cite{chi2-in-lattice} and $\chi ^{(3)}$ \cite{BBB,Jianke,Barcelona}
nonlinearities. Also relevant may be the analysis of the stability
of trapped vortices in a system including competing $\chi ^{(2)}$
and $\chi ^{(3)}$ nonlinearities, cf. the consideration of the
free-space model in work \cite{3Dvortex}.


\begin{thebibliography}{99}

\bibitem{review0} G. I. Stegeman, D. J. Hagan, and L. Torner, ``$\chi ^{(2)}$
cascading phenomena and their applications to all-optical signal processing,
mode-locking, pulse compression and solitons," Opt. Quant. Electr. \textbf{28%
}, 1691-1740 (1996).

\bibitem{review1} C. Etrich, F. Lederer, B. A. Malomed, T. Peschel, and U.
Peschel, ``Optical solitons in media with a quadratic nonlinearity,"
Progr. Opt. \textbf{41}, 483-568 (2000).

\bibitem{review2} A. V. Buryak, P. Di Trapani, D. V. Skryabin, and S.
Trillo, ``Optical solitons due to quadratic nonlinearities: from
basic physics to futuristic applications," Phys. Rep. \textbf{370},
63-235 (2002).

\bibitem{review} B. A. Malomed, D. Mihalache, F. Wise, and L. Torner,
``Spatiotemporal optical solitons," J. Optics B: Quant. Semicl. Opt. \textbf{%
7}, R53-R72 (2005).

\bibitem{KA} Y. S. Kivshar and G. P. Agrawal, \textit{Optical Solitons: From
Fibers to Photonic Crystals} (Academic, 2003).

\bibitem{Rubenchik} A. A. Kanashov and A. M. Rubenchik, ``On
diffraction and dispersion effect on three wave interaction,"
Physica D \textbf{4}, 122-134 (1981).

\bibitem{Romania} D. Mihalache, ``Linear and nonlinear light
bullets: Recent theoretical and experimental studies," Rom. J. Phys. \textbf{%
57}, 352-371 (2012).

\bibitem{first-exper} W. E. Torruellas, Z. Wang, D. J. Hagan, E. W.
VanStryland, G. I. Stegeman, L. Torner, and C. R. Menyuk,
``Observation of two-dimensional spatial solitary waves in a
quadratic medium," Phys. Rev. Lett. \textbf{74}, 5036-5039 (1995).

\bibitem{Drummond} B. A. Malomed, P. Drummond, H. He, A. Berntson, D.
Anderson, and M. Lisak, ``Spatiotemporal solitons in
multidimensional optical media with a quadratic nonlinearity," Phys.
Rev. E \textbf{56}, 4725-4735 (1997).

\bibitem{add1} D. V. Skryabin and W. J. Firth, ``Generation
and stability of optical bullets in quadratic nonlinear media," Opt.
Commun. \textbf{148}, 79-84 (1998).

\bibitem{add2} D. Mihalache, D. Mazilu, J. Dorring, and L. Torner,
``Elliptical light bullets," Opt. Commun. \textbf{159}, 129-138
(1999).

\bibitem{add3} D. Mihalache, D. Mazilu, L.-C. Crasovan, L. Torner, B. A.
Malomed, and F. Lederer, ``Three-dimensional walking spatiotemporal
solitons in quadratic media," Phys. Rev. E \textbf{62}, 7340-7347
(2000).

\bibitem{Wise1} X. Liu, L. J. Qian, and F. W. Wise, ``Generation of optical
spatiotemporal solitons," Phys. Rev. Lett. \textbf{82}, 4631-4634
(1999).

\bibitem{Wise2} X. Liu, K. Beckwitt, and F. Wise, ``Two-dimensional optical
spatiotemporal solitons in quadratic media,"
Phys. Rev. E \textbf{62}, 1328-1340 (2000).

\bibitem{ultra} H. Leblond and D. Mihalache, ``Models of few
optical cycle solitons beyond the slowly varying envelope
approximation," Phys. Reports \textbf{523}, 61-126 (2013).

\bibitem{half} F. A. Bovino, M. Braccini, and C. Sibilia, ``Orbital angular
momentum in noncollinear second-harmonic generation by off-axis
vortex beams," J. Opt. Soc. A, B \textbf{28}, 2806-2811 (201si1).

\bibitem{splitting1} W. J. Firth and D. V. Skryabin, ``Optical solitons
carrying orbital angular momentum," Phys. Rev. Lett. \textbf{79},
2450-2453 (1997).

\bibitem{splitting2} L. Torner and D. V. Petrov, ``Azimuthal instabilities
and self-breaking of beams into sets of solitons in bulk
second-harmonic generation," Electron. Lett. \textbf{33}, 608-610
(1997).

\bibitem{splitting4} D. V. Skryabin and W. J. Firth, ``Instabilities of
higher-order parametric solitons: Filamentation versus coalescence,"
Phys. Rev. E \textbf{58}, R1252-R1255 (1998).

\bibitem{splitting5} J. P. Torres, J. M. Soto-Crespo, L. Torner, and D. V.
Petrov, ``Solitary-wave vortices in quadratic nonlinear media," J.
Opt. Soc. Am. B \textbf{15}, 625-627 (1998).

\bibitem{splitting-exp} D. V. Petrov, L. Torner, J. Martorell, R. Vilaseca,
J. P. Torres, and C. Cojocaru, ``Observation of azimuthal
modulational instability and formation of patterns of optical
solitons in a quadratic nonlinear crystal," Opt. Lett. \textbf{23},
1444-1446 (1998).

\bibitem{splitting-3W} J. P. Torres, J. M. Soto-Crespo, L. Torner, and D. V.
Petrov, ``Solitary-wave vortices in type II second-harmonic
generation," Opt. Commun. \textbf{149}, 77-83 (1998).

\bibitem{splitting-3W2} G. Molina-Terriza, E. M. Wright, and L. Torner,
``Propagation and control of noncanonical optical vortices," Opt.
Lett. \textbf{26}, 163-165 (2001).

\bibitem{Kruglov} V. I. Kruglov, Y. A. Logvin, and V. M. Volkov, ``The
theory of spiral laser-beams in nonlinear media," J. Mod. Opt.
\textbf{39}, 2277-2291 (1992).

\bibitem{BEC} C. J. Pethick and H. Smith, \textit{Bose-Einstein condensate
in dilute gas }(Cambridge University Press, 2008).

\bibitem{cubic-in-trap1} F. Dalfovo and S. Stringari, ``Bosons in
anisotropic traps: Ground state and vortices," Phys. Rev. A
\textbf{53}, 2477-2485 (1996).

\bibitem{cubic-in-trap2} R. J. Dodd, ``Approximate solutions of the
nonlinear Schr\"{o}dinger equation for ground and excited states of
Bose-Einstein condensates," J. Res. Natl. Inst. Stand. Technol. \textbf{101}%
, 545-552 (1996).

\bibitem{cubic-in-trap3} T. J. Alexander and L. Berg\'{e}, ``Ground states
and vortices of matter-wave condensates and optical guided waves,"
Phys. Rev. E \textbf{65}, 026611 (2002).

\bibitem{cubic-in-trap4} L. D. Carr and C. W. Clark, ``Vortices in
attractive Bose-Einstein condensates in two dimensions," Phys. Rev.
Lett. \textbf{97}, 010403 (2006).

\bibitem{cubic-in-trap5} D. Mihalache, D. Mazilu, B. A. Malomed, and F.
Lederer, ``Vortex stability in nearly-two-dimensional Bose-Einstein
condensates with attraction," Phys. Rev. A \textbf{73}, 043615
(2006).

\bibitem{cubic-in-trap6} L. D. Carr and C. W. Clark, ``Vortices and ring
solitons in Bose-Einstein condensates," Phys. Rev. A \textbf{74},
043613 (2006).

\bibitem{cubic-in-trap7} G. Herring, L. D. Carr, R. Carretero-Gonz\'{a}lez,
P. G. Kevrekidis, and D. J. Frantzeskakis, ``Radially symmetric
nonlinear states of harmonically trapped Bose-Einstein condensates,"
Phys. Rev. A \textbf{77}, 043607 (2008).

\bibitem{we} H. Sakaguchi and B. A. Malomed, ``Stabilizing single- and
two-color vortex beams in quadratic media by a trapping potential,"
J. Opt. Soc. Am. B \textbf{29}, 2741-2748 (2012).

\bibitem{Du} F. Du, Y. W. Lu, and S. T. Wu, ``Electrically tunable
liquid-crystal photonic crystal fiber," Appl. Phys. Lett.
\textbf{85}, 2181-2183 (2004).

\bibitem{Luan} F. Luan, A. K. George, T. D. Hedeley, G. J. Pearce, D. M.
Bird, J. C. Knight, and P. S. J. Russell, ``All-solid photonic bandgap
fiber," Opt. Lett. \textbf{29}, 2369-2371 (2004).

\bibitem{BEC0} P. D. Drummond, K. V. Kheruntsyan, and H. He, ``Coherent
molecular solitons in Bose-Einstein condensates," Phys. Rev. Lett. \textbf{81%
}, 3055-3058 (1998).

\bibitem{BEC1} D. J. Heinzen, R. Wynar, P. D. Drummond, and K. V.
Kheruntsyan, ``Superchemistry: dynamics of coupled atomic and
molecular Bose-Einstein condensates," Phys. Rev. Lett. \textbf{84},
5029-5033 (2000).

\bibitem{BEC2} J. J. Hope and M. K. Olsen, ``Quantum superchemistry:
Dynamical quantum effects in coupled atomic and molecular
Bose-Einstein condensates," Phys. Rev. Lett. \textbf{86}, 3220-3223
(2001).

\bibitem{BEC4} T. Hornung, S. Gordienko, R. de Vivie-Riedle, and B. J.
Verhaar, ``Optimal conversion of an atomic to a molecular
Bose-Einstein condensate," Phys. Rev. A \textbf{66}, 043607 (2002).

\bibitem{Tosi} D. L. Feder, C. W. Clark, and B. I. Schneider, ``Vortex
stability of interacting Bose-Einstein condensates confined in
anisotropic harmonic traps," Phys. Rev. Lett. \textbf{82}, 4956
(1999).

\bibitem{Isaac} I. N. Towers, B. A. Malomed, and F. W. Wise, ``Light bullets
in quadratic media with normal dispersion at the second harmonic,"
Phys. Rev. Lett. \textbf{90}, 123902 (2003).

\bibitem{chi2-in-lattice} Z. Y. Xu, Y. V. Kartashov, L. C. Crasovan, D.
Mihalache, and L. Torner, ``Multicolor vortex solitons in
two-dimensional photonic lattices," Phys. Rev. E \textbf{71}, 016616
(2005).

\bibitem{BBB} B. B. Baizakov, B. A. Malomed, and M. Salerno,
``Multidimensional solitons in periodic potentials". Europhys. Lett. \textbf{%
63}, 642-648 (2003).

\bibitem{Jianke} J. Yang and Z. H. Musslimani, ``Fundamental
and vortex solitons in a two-dimensional optical lattice," Opt.
Lett. \textbf{28}, 2094-2096 (2003).

\bibitem{Barcelona} D. Mihalache, D. Mazilu, F. Lederer, Y. V. Kartashov,
L.-C. Crasovan, and L. Torner, ``Stable three-dimensional
spatiotemporal solitons in a two-dimensional photonic lattice,"
Phys. Rev. E \textbf{70}, 055603 (2004).

\bibitem{3Dvortex} D. Mihalache, D. Mazilu, L. C. Crasovan, I. Towers, B. A.
Malomed, A. V. Buryak, L. Torner, and F. Lederer, ``Stable
three-dimensional spinning optical solitons supported by competing
quadratic and cubic nonlinearities," Phys. Rev. E \textbf{66},
016613 (2002).
\end{thebibliography}
\end{document}